\documentclass[a4paper,fleqn]{cas-sc}

\usepackage[authoryear]{natbib}
\usepackage{graphicx} 
\usepackage{float}
\usepackage{algorithm}  
\usepackage{algpseudocode}
\usepackage{color}
\usepackage{setspace}
\usepackage[nomarkers,figuresonly]{endfloat}
\usepackage{units}

\def\tsc#1{\csdef{#1}{\textsc{\lowercase{#1}}\xspace}}
\tsc{WGM}
\tsc{QE}
\tsc{EP}
\tsc{PMS}
\tsc{BEC}
\tsc{DE}
\usepackage[mathlines]{lineno}
\usepackage{etoolbox}
\newcommand*\linenomathpatch[1]{%
  \cspreto{#1}{\linenomath}%
  \cspreto{#1*}{\linenomath}%
  \csappto{end#1}{\endlinenomath}%
  \csappto{end#1*}{\endlinenomath}%
}
\newcommand*\linenomathpatchAMS[1]{%
  \cspreto{#1}{\linenomathAMS}%
  \cspreto{#1*}{\linenomathAMS}%
  \csappto{end#1}{\endlinenomath}%
  \csappto{end#1*}{\endlinenomath}%
}

\expandafter\ifx\linenomath\linenomathWithnumbers
  \let\linenomathAMS\linenomathWithnumbers
  \patchcmd\linenomathAMS{\advance\postdisplaypenalty\linenopenalty}{}{}{}
\else
  \let\linenomathAMS\linenomathNonumbers
\fi
\linenomathpatch{equation}
\linenomathpatchAMS{gather}
\linenomathpatchAMS{multline}
\linenomathpatchAMS{align}
\linenomathpatchAMS{alignat}
\linenomathpatchAMS{flalign}

\begin{document}
\let\WriteBookmarks\relax
\def\floatpagepagefraction{1}
\def\textpagefraction{.001}
\shorttitle{Simple El Ni\~{n}o prediction}
\shortauthors{Sugiura and Kouketsu}

\title [mode = title]{Simple El Ni\~{n}o prediction scheme using the signature of climate time series}

\author[1]{Nozomi Sugiura}[type=editor,
                        auid=000,bioid=1,orcid=0000-0001-6634-1340]
\credit{Conceptualization, Methodology, Software, Writing- Original draft preparation.}

\author[1]{Shinya Kouketsu} 
\credit{Investigation, Writing- Reviewing and Editing.}

%\author[3]{Author 3}
%\credit{Author 3 contribution}

\address[1]{Research Institute for Global Change, Japan Agency for Marine-Earth Science and Technology, Yokosuka, Japan}
%\address[2]{Author 2 affiliation}
%\address[3]{Author 3 affiliation} 

\begin{abstract}
		El Ni\~{n}o is a typical example of a coupled atmosphere--ocean phenomenon,
		but it is unclear whether it can be described quantitatively
		by a correlation between relevant climate events.
		To provide clarity on this issue, we developed a machine learning-based El Ni\~{n}o prediction model that uses the time series of climate indices.
		By transforming the multidimensional time series into the
		path signature,
		the model is able to properly evaluate the order and nonlinearity of
		climate events, which allowed us to achieve good forecasting skill
		(mean square error = 0.596 for 6-month prediction).
		In addition, it is possible
		to provide information
		about the sequence of climate events that tend to change the
		future NINO3.4 sea surface temperatures.
		In forecasting experiments conducted,
		changes in the North Pacific Index and several
		NINO indices were found to be important precursors.
		The results suggest that El Ni\~{n}o
		is predictable to some extent
		based on the correlation of climate events.
\end{abstract}


\maketitle 

\printcredits

\doublespacing

\section{Introduction}
\label{intro}
\label{sec:introduction}
	El Ni\~{n}o is an important climate phenomenon that has
	an immense socio-economic impact. Consequently, its onset/offset mechanism
	has been garnering intense scientific interest
	\citep{Neelin:1998,Wallace:1998,timmermann2018nino},
	and the relationships with the other climate modes and events
	has been investigated \citep{bjerknes1969atmospheric,alexander2002atmospheric,white2014enso} for many years.
	The prediction of El Ni\~{n}o events is still under investigation from various perspectives, including statistical inference from past time series of climate records and results from climate models, initialization from climate models, and data assimilation using oceanic or coupled atmospheric--oceanic models.

	Although previous studies suggest that predictions with expensive climate models
	outperform purely statistical predictions \citep{TwoYearDynamical}, statistical predictions still appear to have value
	because of their simplicity \citep{LIM}.
	Recently, elaborate and quite skillful predictions have been performed 
	based on machine learning, or deep learning,
	with the use of past oceanic sea surface temperatures (SSTs) and subsurface
	information \citep{Wang2020,DEEP}.
	\cite{hu2021benchmarking} extensively evaluated 
	the climate network method that utilizes the relationship
	between spatio-temporal points and concluded that 
	it has some prediction skill over one year.
        \cite{dijkstra2019application} also reported that machine learning models
	can improve the prediction skill over one year.
	Nonetheless, 
	few practical prediction studies have
	employed only the series of multidimensional climate indices as learning datasets.
	As a remarkable exception, \cite{yan2020temporal}
	used the NINO3.4 and Southern Oscillation indices exclusively
	and successfully performed a skillful prediction of the NINO3.4 index
	via a temporal convolutional network.
	In light of the well-known fact that the climate events
	and variabilities in the extratropics can also  modulate 
        El Ni\~{n}o and the Southern Oscillation (ENSO) variability
	\citep{vimont2003seasonal,nakamura2006influence},
	the use of the various climate indices, which represent
	the typical patterns of the ocean surface variables
	(e.g., sea surface temperature and sea level pressure)
	in the course of climate changes, can simply clarify
	the relationships between ENSO and the other climate modes
	as well as improve the prediction efficiency of ENSO.
	However, a possible disadvantage of statistical predictions
	is that
	they typically provide little information about the correlations
	among these climate events through their evolution.
	
	To rectify this issue, we propose a new statistical method that is simple, considerably skillful,
	and provides process information to explain how climate events evolved.
	This study was conducted to develop a practical machine learning-based El Ni\~{n}o prediction scheme using the past time series of climate indices.
	The key ingredient that enables the faithful interpretation of
	the past time series, including their nonlinearities,
	is the signature of paths,
	which is a central concept in rough path theory \citep{lyons2007differential}.
	
	Although several studies have been published on the methodology of
	time series analysis using the signature method \citep{morrill2021generalised},
	there appears to be
	no application of the method to global-scale climate events,
	and thus this paper opens a new field of research in geosciences.
\section{Methodology}
	In this study, we apply supervised learning to a time series of past climate indices.
	We utilize the fact that each segment of the time series is an explanatory variable that is equipped with future values at that time, which can be regarded as objective variables.
	The most significant aspect of the proposed method is that each segment of the time series is transformed into a signature.
	Therefore, our case study simply employs the simplest setting,
	utilizing the signature method to concentrate on the proof-of-concept.
	In this section, after explaining the theoretical basis of why the signatures are relevant, 
	we present the machine learning procedure based on that theory.
	Then, we discuss how to interpret the results and,
	finally, we present the parameters used.
	
	\subsection{Approximating a function on a set of paths\label{iis}}
	In prediction study based on the learning of time series,
	it is crucial to properly construct a predictor
	that link past time series segments to future values.
	A predictor is represented as a continuous function on a set of multidimensional paths.
	To secure the performance of the predictor,
	it is essential to choose an appropriate basis for the function of the path
	because it determines the expressive power of the function.
	Note that our concern is not the basis for a path but for a function of paths.
	In this sense, the most mathematically justified candidate for the basis
	is the signature \citep{lyons2007differential,Sugiura2020}.

	For a $d$-dimensional path $X=X_{[s,t]}: [s,t] \to \mathbb{R}^d$ that
	maps $\tau$ to $X_{\tau}$,
	the $0$-th to $n$-th iterated integrals are
	defined recursively, as follows \citep{lyons2007differential}: 
		\begin{align}
			\mathcal{S}^{()}(X_{[s,t]})&=1,\\
			\mathcal{S}^{(i_1\cdots i_n)}(X_{[s,t]})&=\int_s^t
			\mathcal{S}^{(i_1\cdots i_{n-1})}(X_{[s,t_n]})
			dX^{(i_n)}_{t_n},\quad
                        i_1,\cdots, i_n = 1,\cdots, d.
		\end{align}
	The signature $\mathcal{S}(X)$ of path $X$ is the collection of all the iterated integrals,
	and the operation $\mathcal{S}: X\mapsto \mathcal{S}(X)$ is called the signature transform.
	In particular, its truncation up to the $n$-th iterated integrals is called
	the step-$n$ signature, $\mathcal{S}_n(X)
	\in \bigoplus_{k=0}^{n}(\mathbb{R}^d)^{\otimes k}$,
	which means that the multi-index $I$ in
	component $\mathcal{S}_n^{(I)}(X_{[s,t]})$ runs across
		\begin{align}
                  I &\in \phi \cup   \{1,\cdots,d\} \cup \cdots
                  \cup \{1,\cdots,d\}^n.
		\end{align}
	
	Now, let $C(K,\mathbb{R})$ be the space of a continuous function on
        a compact set $K$ of paths with at least one monotonous coordinate.
        Subset $A \subset C(K,\mathbb{R})$ is defined as
        {\small
		\begin{align}
			A &=
                        \left\{g: X \mapsto \sum_I w^{(I)} \mathcal{S}^{(I)}(X) 
			\middle| 
			w \in \bigoplus_{k=0}^{n}(\mathbb{R}^d)^{\otimes k}
			\text{~for some $n\geq 0$}\right\}.
		\end{align}
                }
	Then, $A$ satisfies the following conditions:
	\begin{enumerate}
		\item Because the 
		step-$n$ signature transform $K\ni X\mapsto \mathcal{S}_n(X)$ is continuous
		for any $n>0$, $A\subset C(K,\mathbb{R})$.
		\item $g_1,g_2 \in A \text{ and } \lambda_1, \lambda_2 \in \mathbb{R}
		\implies \lambda_1 g_1 + \lambda_2 g_2\in A$.
		\item Constant-valued function $\mathbf{1} \in A$.
		\item Based on the shuffle identity \citep{lyons2007differential},
		$g_1,g_2 \in A \implies g_1g_2 \in A$.
		\item Based on the uniqueness theorem \citep{levin2013learning},
		for all $X,Y\in A$ with $X\neq Y$,
		there exists $g\in A$ that satisfies $g(X)\neq g(Y)$.
	\end{enumerate}
	From these conditions, we can apply the
	Stone--Weierstrass theorem \citep{Stone1937ApplicationsOT} to the subset $A$
	and conclude that 
	$A$ is dense in $C(K,\mathbb{R})$,
	which means that any function $f\in C(K,\mathbb{R})$
	is uniformly approximated by a function $g\in A$ with
        arbitrary accuracy \citep{levin2013learning,fermanian2021embedding}.
	
	From the above reasoning, we can construct
	a nonlinear predictor as 
	a linear combination of iterated integrals for
	each segment of the multidimensional time series.
	
	\subsection{Procedure for machine learning of time series}
	In the proposed approach,
	the predictor is constructed as follows.
	Suppose we have a time series of several climate indices
	defined at each calendar month of $t_m,~m=1,2,\cdots,M$.
	Any segment of the time series over a period, say six months,
	can be viewed as a multidimensional path, which can be represented by
	the signature.
	
	For this supervised learning,
	the object variable is the NINO3.4 index $y_{m+m_a}$ at time $\tau=t_{m+m_a}$,
	while the explanatory variables  $x_m$ are the signature 
	for the segment of time series $X$ in the period $[t_{m-m_b+1},t_m]$. 
	The approximation property described in the previous section
	allows us to express the object variable as a linear combination
	of the explanatory variables:
		\begin{align}
			y_{m+m_a} &= y_{m}+\left< w_m, x_{m}\right>+\epsilon,\\
			x_{m} &:=
			\mathcal{S}_n(X_{[t_{m-m_b+1},t_m]}),\label{sig_model}
		\end{align}
	where $\mathcal{S}_n(X_{[t_0,t_1]})$ denotes the order-$n$ signature for the $d$-dimensional
	time series in the interval $[t_0,t_1]$,
	$\left<a,b\right>$ denotes the scalar product $\sum_I a^{(I)}b^{(I)}$,
	$w_m=\{w_m^{(I)}|I=\text{multi-index}\}$ is the weight vector for the predictor, 
	$\epsilon$ is a random variable representing prediction error,
	$t_m$ is the starting time of the prediction, $t_{m-m_b+1}$ is
	the starting point of the path segment,
	and $t_{m+m_a}$ is the target time for prediction.
	\begin{figure}
\centering
			\includegraphics[width=0.9\textwidth,clip]{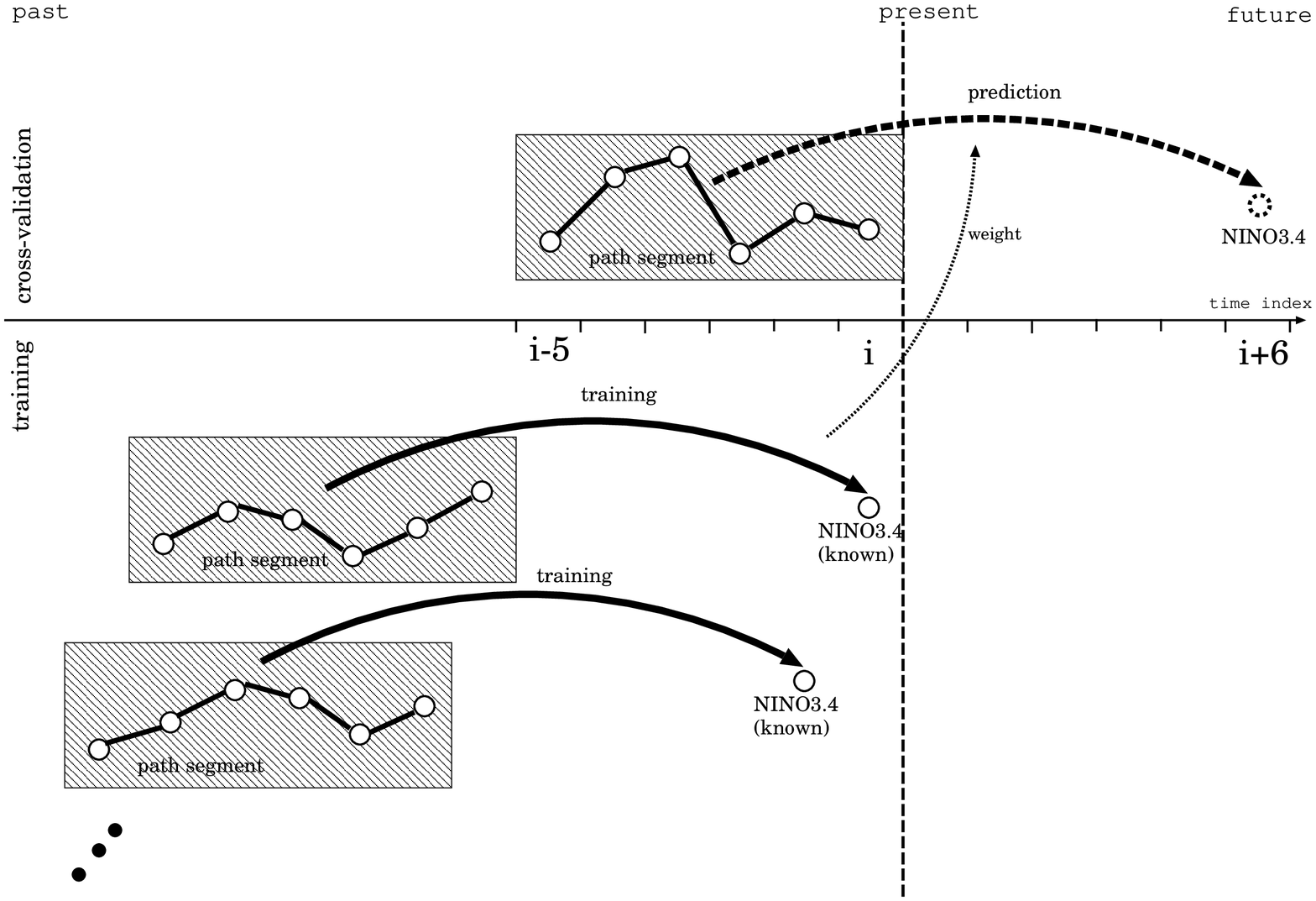}
		\caption{Schematic view of training and prediction flow, assuming that $m_b$=$m_a$=6.
			Hatched squares represent transforming into the signature.
			Predicted value at time index $m$+6 will be compared with the validation data if available.
			\label{flow}}
	\end{figure}
	Before converting into the signature,
	a zero vector is added at the beginning of each series $X_{[t_{m-m_b+1},t_m]}$
	to account for the magnitude of the value at the starting point
	\citep{morrill2021generalised}.
	We computed the signature by using the 
	Python library ${\tt esig}$ \citep{esig}.
	
	In the control case, we instead used the time series as is without converting it
	into the signature:
		\begin{align}
			x_{m} &:= X_{[t_{m-m_b+1},t_m]}=
			\left(
			X_{t_{m-m_b+1}},
			X_{t_{m-m_b+2}},\cdots,
			X_{t_{m}}
			\right).\label{ctl_model}
		\end{align}
	This corresponds to an auto-regressive (AR) model.
	
	Using the training dataset available up to time $t_m$,
		\begin{align}
			D_m &=
			\left\{\left(x_{\mu},y_{\mu+m_a}-y_{\mu}\right)
			\middle|~ \mu \in [m_b,m-m_a]
			\right\},
		\end{align}
	we first estimate the
	optimal weight $w=w_{m}$ that minimizes the cost function
	with an $L_1$-penalty term:
		\begin{align}
			J_m(w) &= \frac1{2|D_{m}|}  \sum_{\mu =m_b}^{m-m_a}
			\left( y_{\mu+m_a}  - y_{\mu} -\left< w, x_{\mu}\right>\right)^2
			+\alpha \sum_I\left| w^{(I)} \right|,
		\end{align}
	where $|D_{m}|=m-m_a-m_b+1$ is the number of samples in $D_{m}$, and $I$ is the multi-index.
	The optimization problem is solved by the Lasso model fit with least angle regression
	\citep{scikit-learn}, which is suitable for problems with many parameters.
	We then predict
	a future NINO3.4 index as
	$\widehat{y}_{m+m_a} =
	y_m + \left< w_m, x_{m}\right>$ and compare it to $y_{m+m_a}$.
	In other words, a cross-validation is made against the validation data:
		\begin{align}
			D'_m &=
			\left\{\left(x_{m},y_{m+m_a}-y_{m}\right)
			\right\}.
		\end{align}
	We repeat the above procedure after incrementing the time index $m$ by $1$.
	
	Figures\,\ref{flow} shows the schematic view of training and prediction flow.
	In this flow, 
	the weight $w_m$ is obtained by using the training dataset $D_m$, and
	then the prediction from time $t_m$ using the signature $x_m$ yields the value 
	$\widehat{y}_{m+m_a}$, which is subject to comparison with the validation data
	$y_{m+m_a}$. 
	Note that the size of the training data $|D_m|=m-m_a-m_b+1$
	depends on the starting time $t_m$.
	The prediction error can be obtained
	from the statistics of $\widehat{y}_{m+m_a}-y_{m+m_a}$ for various starting times.
	By taking this approach, where training and forecasts are done progressively
	by moving the starting time of the forecast hiding future at that moment,
	each forecast is assured to be a fair cross-validation.

	\subsection{Diagnosing the dominant event sequences}
	One difficulty with regular machine learning is that it does not provide
	sufficient reasoning for the results.
	However, the signature-based method allows us to mathematically extract
	from the path 
	those properties that are important in the prediction.
	
	To diagnose the dominant event sequences 
	that contribute to the prediction,
	we compute 
	the standard partial regression coefficients (SPRCs)
	$r_{m}^{(I)}$,
	which represent the sensitivity of normalized value $y_{\mu+m_a}$
	in the future to each component of the 
	normalized signature $x_{\mu}^{(I)}$ in the past, as 
		\begin{align}
			r_{m}^{(I)}
			&= \frac{\sigma_{x_m^{(I)}}}{\sigma_{y_m}}w_{m}^{(I)},\label{sprc}
		\end{align}
	where $\sigma_{y_m}$ and $\sigma_{x_m^{(I)}}$ denote the standard deviations
	of $y_{\mu+m_a}-y_{\mu}$ and $x_{\mu}^{(I)}$, respectively, 
	among the samples in $D_{m}$, which represents 
	the learning data in the period from time $t_{m_b}$ to time $t_{m-m_a}$.
	
	\subsection{Setting of experimental parameters}
	We used a climate time series composed of $d=12$ indices in Table\,\ref{indices}
	retrieved from NOAA cite \citep{PSL}.
	The time series starts at $t_1$ (January of 1900),
	and ends at $t_{1459}$ (July of 2021).
        
	\begin{table}
		\begin{center}
			\caption{Twelve climate indices and their abbreviations\label{indices}}
			\begin{tabular}{l|l|l}
				Abbrev. & Climate Indices  & References\\
				\hline 
				NINO34  &  Nino 3.4 (5N-5S, 170W-120W) SST &\cite{nino12}\\
				NINO12  & Nino 1+2 (0-10S, 90W-80W) SST &\cite{nino12}\\
				NINO3   & Nino 3 (5N-5S, 150W-90W) SST &\cite{nino12}\\
				NINO4   & Nino 4 (5N-5S, 160E-150W) SST &\cite{nino12}\\
				DMI     & Dipole Mode Index &\cite{DMI}\\
				AMO     & Atlantic Multidecadal Oscillation index &\cite{AMO}\\
				NPI     & North Pacific Index &\cite{NPI}\\
				SOI     & Southern Oscillation Index &\cite{ropelewski1987extension}\\
				NAO     & North Atlantic Oscillation (NAO) index &\cite{NAO}\\
				TPI     & Tripole Index &\cite{2015ClD}\\
				AO     & Arctic Oscillation index &\cite{AO}\\
				MON & Date elapsed (mid-day in month divided by 365)&-\\
				\hline
			\end{tabular}
		\end{center}
	\end{table}
	
	The standard lead time for prediction is $6$ months ($m_a=6$),
	whereas each past segment is of length $6$ months ($m_b=6$).
	The experiment duration was from the prediction starting at $t_{961}$ (January of 1980),
	to the one starting at $t_{1453}$ (January of 2021).
	
	We used iterated integrals up to level $n=3$, which means that the 
	total number of terms in the linear combination was $N=(d^{n+1}-1)/(d-1)=1885$,
	and the intensity of the $L_1$ penalty term was tuned 
	to $\alpha=2.0$.

\section{Results}
\begin{figure}
          \centering
		\includegraphics[width=0.9\textwidth,clip]{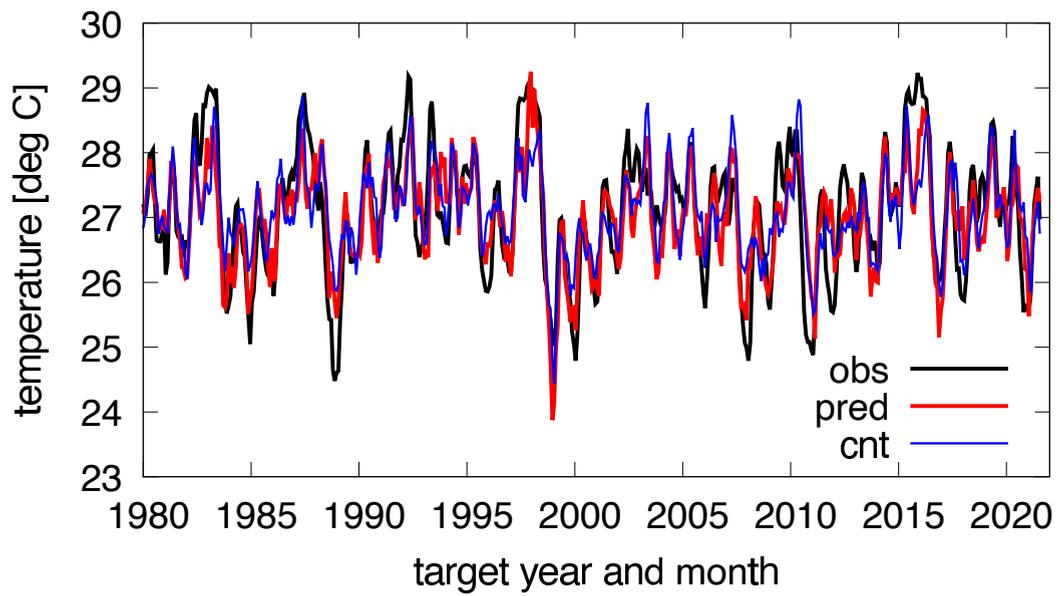}
		\caption{Comparison of NINO3.4 for 6-month predictions.
			Red: signature case; blue: control case.
			Horizontal axis is the target year and month,
			and vertical axis is temperature in $\unit{^{\circ} C}$.
			\label{pred_result0}}
	\end{figure}

\begin{figure}
          \centering
	 	\includegraphics[width=0.9\textwidth,clip]{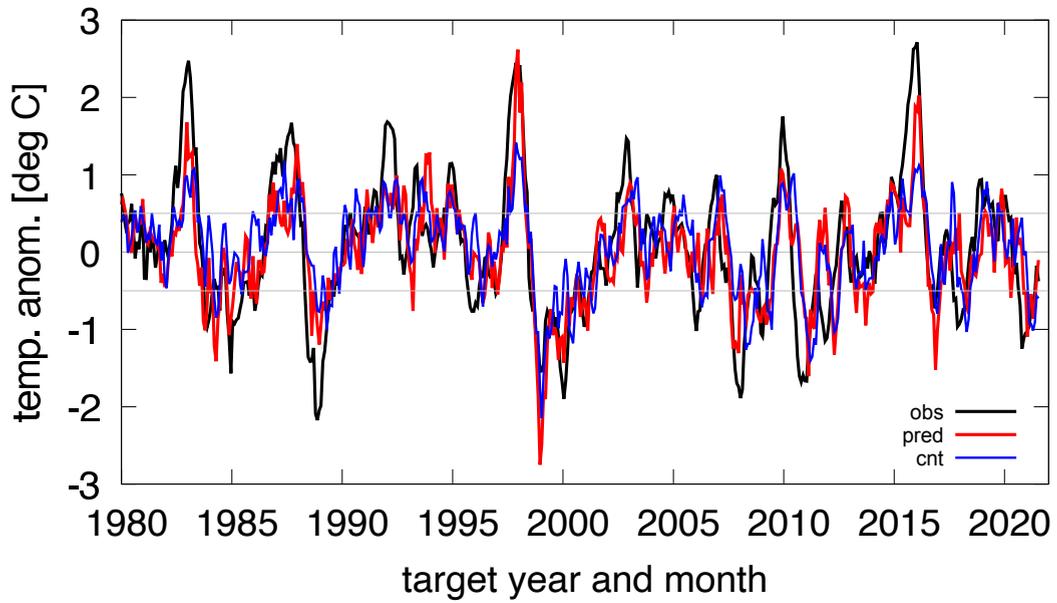}
	 	\caption{The same as Fig.\,\ref{pred_result0} but shown
                   as anomalies, which are defined as
                   the difference from the past 30-yr mean of
                   monthly values.
	 		Red: signature case; blue: control case.
	 		Horizontal axis is the target year and month,
	 		and vertical axis is temperature anomalies  in $\unit{^{\circ} C}$.
	 		\label{pred_result}}
	 \end{figure}

\begin{figure}
          \centering
		\includegraphics[width=0.9\textwidth,clip]{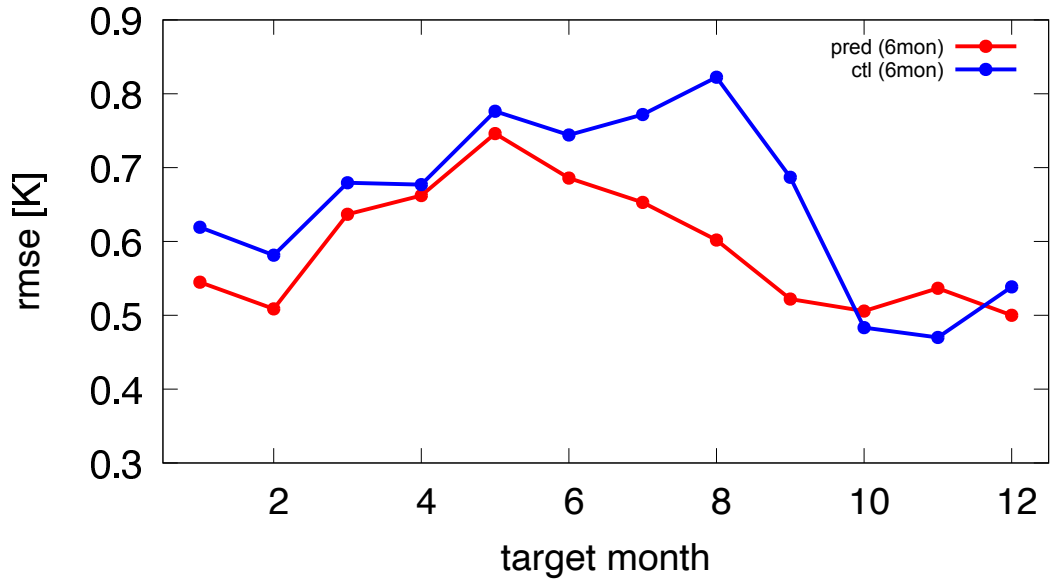}
		\caption{Prediction error for each target month.
			Red: signature case; blue: control case.
			Horizontal axis is the target month (1 = January, 2 = February,
			$\cdots$, 12 = December),
			and vertical axis is rms error in $\unit{K}$.
			\label{rmse_result}}
	\end{figure}

	Figure\,\ref{pred_result0} shows the result of $6$-month prediction,
	whereas fig\,\ref{pred_result} shows it as anomalies from climatology,
        These figures indicate the overall superiority of prediction using the signature.
	The prediction error for each target month is shown in Fig.\,\ref{rmse_result}.
	It is obvious that the predictions for July to September were much better than those in
	the control case; however, they were comparable in the other months.
	The overall prediction skill was $0.596\unit{K}$
	for the signature case and $0.663\unit{K}$ for the control case.
	
	\cite{Wang2020} proposed 
	an operator-theoretic technique called kernel analog forecasting (KAF),
	which has
	a rigorous connection with Koopman operator theory for dynamical systems,
	yielding statistically optimal predictions as conditional expectations.
	They also compared it to the linear inverse model (LIM),
	which is a linear evolution operator for modes.
	Note that both methods employ as explanatory variables
	the dominant modes in spatiotemporal SST,
	which we did not use in our study.
	Here, we use KAF and LIM for the comparison of forecasting skill.
	For the comparison with KAF and LIM,
	the root-mean square (rms) errors for 6-month prediction 
	in the period from 1998 to 2017 were computed.
	The signature model, AR model, KAF model, and LIM
	had rms values of  $0.617$, $0.686$, $0.62$, and $0.75\unit{K}$, respectively.
	This comparison result suggests that the forecasting skill of the signature model is comparable to that of the KAF model.
	
	The spring prediction barrier is defined in \cite{lai2018enso} as follows:
	``... models have problems in predicting Boreal winter tropical
	Pacific sea surface temperature (SST) when forecasts start in
	Boreal spring (February–May).
	This is called the spring predictability barrier. ''
	Similarly,
	\cite{zheng2010spring} pointed out that 
	``... errors have the largest values and the fastest growth rates initialized before and during the NH spring.''
	In light of these definitions,
	the spring predictability barrier, i.e., 
	poor prediction skill when starting from February and March,
	seems to disappear as indicated by the rms error values in the
	target months of August to September. 
	
\begin{figure}
		\begin{center}
			\includegraphics[width=0.8\textwidth,clip]{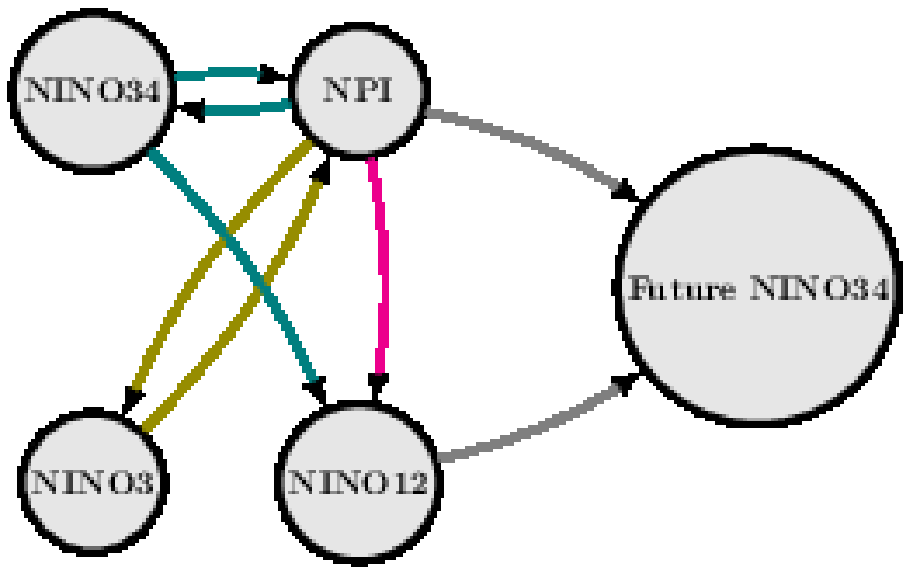}
			\caption{Typical climate event flows for predicting future Nino3.4 index.
				Arrows indicate time order.
				Key indices include 
				NINO12 (Nino1+2 SST), NINO34 (Nino3.4 SST),
				NINO4 (Nino4 SST), and NPI (North Pacific Index).
				\label{events}}
		\end{center}
	\end{figure}

	Table\,\ref{dominant} shows the dominant event sequences 
	among iterated integrals.
	The events with the first to the third indices are shown in each row. 
	If the same index appears twice in a row, then the event is intense.
	The top sequence in the period from 1900 to 2020 is an intense NPI change followed by a Ni\~{n}o 1+2 SST change.
	The key indices are NPI and various NINO indices.
	In particular, 
	NPI, an atmospheric process, is involved in all the dominant sequences,
	which should be a manifestation that El Ni\~{n}o is a coupled atmospheric--oceanic process.
	In addition, the comparison between statistics for two different periods suggests that the Nino1+2 index,
	corresponding to the region of coastal South America,
	is becoming more important as  a precursor in the 21st century.
	Summarizing the above,
	fig.\,\ref{events} illustrates how the dominant climate events occur
	that will lead to changes in the future NINO3.4.
	
	Although the main indices in terms of the iterative integrals
	are related to water temperature in the NINO regions,
	it appears that not only these but also various climate indices contribute
	incrementally, and the predictor is built on the balance of them.
 In fact, if we perform an experiment with the main 5 indices (NINO12, NINO3, NINO34, NINO4, and NPI),
 the rms error for the prediction is $0.666\unit{K}$, with no improvement from the control case.
	In this respect, the inference structure is considered to be different
	from other SST-based predictions such as KAF and LIM.
	%199907
	% (4,9,4): NPI-NINO3-NPI  5.55
	% (9,4,4): NINO3-NPI-NPI -3.70
	% (4,4,5): NPI-NPI-NINO12 3.63
	% (2,2,4): NINO34-NINO34-NPI -3.57
	% (2,4,4): NINO34-NPI-NPI 3.34
	%202102
	%(4,4,5): NPI-NPI-NINO12 4.53
	%(2,2,4): NINO34-NINO34-NPI -4.17
	%(4,2,5): NPI-NINO34-NINO12 -3.37
	%(2,4,5): NINO34-NPI-NINO12 -3.25
	%(4,9,4): NPI-NINO3-NPI 2.88
	\begin{table}
		\begin{center}
			\caption{Top five dominant event sequences among iterated integrals.
				“1st” denotes the first index for the corresponding 
				iterated integral:
				$\mathcal{S}^{(i_1i_2i_3)}(X)=\int_s^t\int_s^{t_3}\int_s^{t_2}
				dX_{t_1}^{(i_1)}dX_{t_2}^{(i_2)}dX_{t_3}^{(i_3)}$.
				Events happen from first to third: $t_1 < t_2 < t_3$.
				If the same index appears twice in a row, then the event is intense.
				``SPRC'' represents the standard partial regression coefficients (Eq.\,\ref{sprc}).
				\label{dominant}}
			\begin{tabular}{r|r|lll}
				& \multicolumn{4}{c}{Learning data from Jan. 1900 to Dec. 1999}\\
				& SPRC & \multicolumn{3}{c}{Indices in iterated integrals}\\
				No.&$r_m^{(i_1i_2i_3)}$
				&1st ($i_1$)& 2nd ($i_2$)&3rd ($i_3$)\\
				\hline
				1& $5.55$  &NPI & NINO3 & NPI \\
				2& $-3.70$  &NINO3 & NPI & NPI \\
				3& $3.63$  &NPI & NPI & NINO12 \\
				4& $-3.57$ &NINO34 &NINO34 & NINO34 \\
				5& $3.34$  &NINO34 & NPI & NPI \\
				\hline
				\hline
				& \multicolumn{4}{c}{Learning data from Jan. 1900 to Dec. 2020} \\
				& SPRC & \multicolumn{3}{c}{Indices in iterated integrals}\\
				No.&$r_m^{(i_1i_2i_3)}$
				&1st ($i_1$)& 2nd ($i_2$)&3rd ($i_3$)\\
				\hline
				1& $4.53$  &NPI & NPI & NINO12 \\
				2& $-4.17$  &NINO34 & NINO34 & NPI  \\
				3& $-3.37$  &NPI & NINO34 & NINO12 \\
				4&  $-3.25$  &NINO34 & NPI & NINO12 \\
				5&  $2.88$  &NPI &  NINO3 &NPI\\
				\hline
			\end{tabular}
		\end{center}
	\end{table}
	
	
\section{Conclusions}
	We developed a model that can statistically predict
	El Ni\~{n}o using only the time series of past multidimensional climate indices.
	By converting the time series into the signature, the accuracy of the machine
	learning algorithm is improved and, thereby, the NINO3.4 SST can be predicted
	to some extent six months in advance.
	An important byproduct of this approach is that the correlation of 
	climate events can be read from the dominant iterative integral.
	For example, it was suggested that variations in the NPI, NINO12,
	and other indices occur in a certain order,
	which leads to variations in the NINO3.4 SST.
	It was also found that the signature method can learn the nonlinear
	development of El Ni\~{n}o more accurately than the traditional AR model
	and, thus, is less sensitive to the spring barrier of predictability.
	Future research is required to improve the scheme
	by incorporating more detailed oceanographic information,
	evaluating uncertainties, and considering other factors.

	The predictions obtained by this method are not marked with error bars,
	but because we know the prediction error for each month as shown in Fig.\,\ref{rmse_result},
	we can consider these values as the prediction error.
	However, as it is not possible to give a forecast error for each forecast individually,
	an ensemble could be created by bootstrapping or other methods to improve this point,
	which may also lead to a factory for forecast accuracy.
	
	The length of the path segment used for the 6-month forecast, 6 months,
	was confirmed in preliminary experiments (not shown) to be appropriate,
	but the length of the path segment required for forecasts with other lead
	times may be different.
	It is also necessary to confirm whether the step-$3$ signature is optimal.
	
	The dominant iterated integral for prediction
	may change from time to time depending on the period covered,
	as shown in Table\,\ref{dominant}.
	It needs to be carefully considered 
	how this relates to the decadal changes in the
	Ni\~{n}o mechanism.
	These points remain as future work.

\section{Acknowledgments}
	This study was funded by JST-PROJECT-20218919.

\newpage

\textbf{Code availability section}

Name: enso\_signature

Contact: nsugiura@jamstec.go.jp, +81-46-867-9054

Hardware requirements: CPU

Program language: Python
 
Software required: Python libraries esig and scikit-learn

Program size: 188 lines

The source codes are available for downloading at the link: 
https://github.com/nozomi-sugiura/enso\_signature

\bibliographystyle{cas-model2-names}
\bibliography{cluster} 

\end{document}